\documentstyle[emulateapj,psfig]{article}
\begin{document}    
\def\etal{{\it et al.\thinspace}}
\newcommand{\be}{\begin{equation}}
\newcommand{\ba}{\begin{eqnarray}}
\newcommand{\ee}{\end{equation}}
\newcommand{\ea}{\end{eqnarray}}  
       
\title{High-Redshift Galaxy Outflows and the Formation of Dwarf Galaxies}

\author{Evan Scannapieco, Robert J. Thacker, \& Marc Davis}
    
\affil{
Department of Astronomy, University of California, Berkeley, CA 94720}

\begin{abstract}

We examine the effects of galaxy outflows on the formation of dwarf
galaxies in numerical simulations of the high-redshift Universe.
Using a Smoothed Particle Hydrodynamic code, we conduct two detailed
simulations of a (5.2 Mpc/$h$)$^3$ comoving volume of the Universe.
In both simulations we implement simple, well-motivated models of
galaxy identification and star formation, while our second simulation
also includes a simple ``blow-out'' model of galaxy outflows in which
supernova driven winds from newly formed disk galaxies punch-out and
shock the intergalactic medium while leaving the host galaxies intact.
A direct comparison between these simulations suggests that there are
two major mechanisms by which outflows affect dwarf formation.
Firstly, the formation of an outflow slows down the further accretion
of gas onto a galaxy, causing an overall decrease of approximately
50\% in the total gas mass accreted by the
objects in our simulations.  Additionally, our simulations uncover a
significant population of $\sim 10^9 M_\odot$ objects whose formation
is suppressed by the mechanism of ``baryonic stripping,'' in which
outflows from early galaxies strip the gas out of nearby overdense
regions that would have otherwise later formed into dwarf galaxies.
This mechanism may be important in explaining the observed discrepancy
between the number of dwarf galaxies predicted and observed in the
local group and provide a natural explanation for the formation of
empty halos which may be required by the existence of the extremely
gas-poor extra-galactic High-Velocity Clouds.

\end{abstract}

\section{Introduction}

It has long been recognized that the properties of the diffuse
gas within clusters of galaxies are indicative of a violent past.
The slope of the X-ray luminosity-temperature relationship of the 
hot intracluster medium  in galaxy clusters
is far too steep to be  due to heating only by viralising shocks,
requiring that large quantities of heated material
be injected into this gas at early times.
(see eg.\ Kaiser 1991; Mushotzky \& Scharf 1997; Eke, Navarro
\& Frenk 1998; Cavaliere, Menci, \& Tozzi 1999).
Similarly, the metallicity of the intracluster medium is observed to be quite
high, $\sim .3 Z_\odot$, and constant over a large range of cluster
masses (Renzini 1997), indicating widespread enrichment by material 
ejected by supernovae.  (For an alternative viewpoint, see however
Bryan 2000 and Pearce et al.\ 2000).

While the first observational evidence for preheating and enrichment 
was met with theoretical resistance, hindsight leads us to
believe that we might have expected this all along.  As 
structure formation is thought to occur hierarchically,
and this growth is biased to the same over-dense regions at all times,
it is natural to expect that a large population of dwarf galaxies
would have once been found near the locations of present-day clusters.
It has also been widely observed that dwarf galaxies
undergo outflows
both at low and high redshift.  Several studies of
expanding HI gas in nearby dwarf galaxies show clear evidence
of dense, expanding shells with velocities above 15 km/s
(Marlowe et al.\ 1995; Heckman 1997; Hunter et al.\ 1998;
Martin 1998) as well as halos of hot ($\sim 5 \times 10^6$ K) 
gas surrounding these objects (della Ceca et al.\ 1996; Bomans, Chu, 
\& Hopp 1997). Similarly, spectroscopic studies of galaxies
have confirmed that high-velocity winds are present around dwarf galaxies 
at redshifts $\sim$ 3 (Pettini et al.\ 1998; Pettini et al.\ 2001)
and even higher (Frye \& Broadhurst 1998; Warren et al.\ 1998).
Taken together these observational facts suggest
a picture in which outflows powered by starbursts
in the earliest dwarf galaxies had a huge impact on the
properties of the gas that later condensed within clusters
of galaxies.

Thus a vague theoretical scheme exists for understanding the
preheating of clusters; yet surprisingly, the impact of this mechanism
on the intergalactic medium (IGM) has been largely unexplored.
Although clusters and galaxies both condense out of the same
material, preheating is only considered in cluster
simulations (see eg.\ Metzler \& Evrard 1994; Yepes at al.\ 1997) and
simulations of the properties of the IGM (Nath \& Trentham 1997;
Madau, Ferrara, \& Rees 2000), while all numerical galaxy formation
simulations carried out to date assume primordial conditions.
Feedback in these simulations is modeled solely as impacting the
interstellar medium  within the galaxy, without affecting the IGM
and neighboring galaxy formation.
 
We have recently conducted 
two exploratory semi-analytical studies on the impact of galaxy outflows
on the IGM and the formation of galaxies within it
(Scannapieco, Ferrara, \& Broadhurst 2000; Scannapieco \& 
Broadhurst 2001, hereafter SB).  These studies have 
shown that outflows may be crucial in resolving such
long-standing astronomical mysteries as the factor of four
discrepancy between the number of observed and predicted
Milky-Way Satellite galaxies (see eg.\ Klypin et al.\ 1999;
Moore et al.\ 1999) and the physical processes
that cause elliptical galaxies to be typically much larger than
disks (see eg.\ Bromley et al.\ 1998).
The results of our semi-analytic simulations hinged on the importance of
a mechanism of ``baryonic stripping'' in which outflows from early
galaxies strip the gas out of nearby overdense regions
of the IGM that would have otherwise later formed into 
dwarf galaxies.

In SB we were able to illustrate the importance of this
mechanism in an idealized context, however the nonlinear clustering of 
galaxies, complicated evolution of outflows, and details of shock
motion through collapsing regions can be captured only through
numerical simulation.  In this article we address this question in the 
more detailed setting of Smoothed Particle Hydrodynamic (SPH)
simulations, studying the 
impact of a simple model of galaxy outflows on the formation of 
structure in a typical volume of the Universe. 


Feedback has been implemented by other authors using a number of
different prescriptions. The lack of a fundamental theory of star
formation and the interstellar medium render it extremely difficult to
decide how to model star formation and feedback `correctly'. For
example, the dispersal of energy from supernovae can be achieved by
thermal heating, by passing kinetic energy to the gas, or by some
arbitrary combination of the two. Katz (1992) was the first to
consider the effect of thermal heating, and this seminal work
highlighted a fundamental problem, namely that the cooling time for
hot dense gas is so short (less than Myr) that simply returning
feedback energy as thermal heating has little effect. Navarro \& White
(1993) considered a feedback model with both kinetic boosts and
thermal heating. Mihos and Hernquist (1994) and Gerritsen (1997) have
studied the effect of feedback modeled as kinetic boosts in isolated
disk galaxies.  Gerritsen has argued that the boost model is too
efficient when energy transfer parameters are chosen to match
theoretical predictions such as those of Thorton et al.\
(1998). This is probably related to the fact that the kinetic energy
boosts are calculated using a fixed energy budget that does not
account for the force softening that occurs at the short scales on
which feedback is modeled.

More recently, Yepes et al.\ (1997) have constructed a multi-phase
model of star formation and feedback, which has been implemented in
SPH simulations by Hultman \& Pharasyn (1999). Given the large number 
of uncertainties in modeling the
evolution of the interstellar medium, it remains unclear if such
sophisticated approaches are well-motivated.  Thacker \& Couchman
(2000) present a number of different approaches to modeling feedback,
including a new approach where drastic radiative losses are prevented
by using a modified cooling formalism.  Springel (2000) has developed
a model where the local evolution of gas can become dominated by
turbulent pressure-support from exploding supernovae. A survey of the
current status of feedback in simulations of galaxy formation can be
found in Thacker \& Davis (1999).

Given the inherent difficulty in modeling feedback, in this work we
adopt a simple kinetic model that is inspired by the numerical
simulations conducted by Mac Low \& Ferrara (1999) and Ferrara \&
Tolstoy (1999).  Here the authors demonstrate that in disk galaxies
within halos of $10^7 M_\odot \lesssim M \lesssim 10^9 M_\odot$, a
``blow-out'' occurs in which super-bubbles formed by groups of Type II
supernovae punch out of the disk, shocking the IGM while leaving the
interstellar medium of the galaxy intact.  This is due to the fact
that while supernovae are inefficient at ejecting material within the
disk, they expand almost freely in the perpendicular direction,
resulting in a large organized wind made up of mostly the gas
surrounding the galaxy.  This scenario has also been supported by
the numerical simulations of Martel \& Shapiro (2000).

The structure of this work is as follows.  In \S2 we outline the numerical
method adopted and the simplifying assumptions used in constructing
our model for galaxy outflows.  In \S3 we compare the results from
hydrodynamic simulations conducted both with and without outflows to 
examine the impact of outflows on the number density and distribution of 
galaxies of varying mass scales.  Conclusions are given in \S4.

\section{Simulating Galaxy Outflows}

\subsection{Overview of the Numerical Method}

We model the gravitational and hydrodynamic forces using a particle-based
method. The gravitational interactions are calculated using the Adaptive
Particle-Particle, Particle-Mesh (AP${}^3$M) algorithm of Couchman (1991),
while hydrodynamic forces are calculated using Smoothed Particle
Hydrodynamics (Gingold \& Monaghan 1977; Lucy 1997). We use a
parallel OpenMP based implementation of the `HYDRA' code (Couchman, Thomas 
\& Pearce 1995; Thacker \& Couchman 2000, in prep) that is 
optimized for execution on RISC processors.

Since we are attempting to calculate ensemble quantities, it is necessary
for us to consider a uniform region of space that is not subject to
bias due to the incorporation of a dominant collapse mode. We must also
ensure that the mass resolution is sufficient to be able to resolve groups
of mass $10^8 M_\odot$, the smallest mass that can cool effectively
without molecular cooling or in the presence of an ionizing
background (see eg.\ Haiman, Rees, \& Loeb 1997; Ciardi et al. 2000; SB).

These criteria, along with those of wall clock
limitations, have lead us to use a periodic box of size $5.2$ $h^{-1}$
Mpc and a particle number of $2\times 192^3$.  The use of such a small
box precludes evolving to low redshift, but this is not a concern since we
are interested in the formation epoch of low ($<10^{10}$ M${}_\odot$)
mass halos. The mass resolution in the dark matter is $2.5 \times 10^6 M_\odot$
while that within the gas is $5 \times 10^5 M_\odot$, 
rendering a $2 \times 10^8$ M${}_\odot$ halo resolved by 80 particles. 
We use a fixed physical Plummer softening length of 1.54 kpc, which leads
to  minimum hydrodynamic scale of $h_{min}=$ 1.8 kpc. 

A detailed discussion of the SPH solver used in this investigation is
given by Thacker et al.\ (2000). Here we discuss features of the algorithm
that intimately relate to the group identification scheme and outflow
processes we are modeling.  

The density at a given point within the simulation is given by
\begin{equation}
<\rho({\bf r}_i)>= \sum_{j=1,r_{ij}<2h_{i}}^N m_j
[W({\bf r}_i-{\bf r}_j,h_i)+W({\bf r}_i-{\bf r}_j,h_j)]/2,
\end{equation}
where $W({\bf r},h)$ is the ($B_2$ spline) SPH smoothing kernel, $h_i$
is the smoothing length associated with particle $i$, $m_j$ is the mass
of particle $j$, and ${\bf r}_i$ and ${\bf r}_j$ are the coordinates
of particles $i$ and $j$, with $r_{ij}$ the distance between them. 
The arithmetic average of
the kernels is used to ensure that the equation of motion is correct to
order $\nabla h$, when used in combination with the gather-only neighbor
finding method (see Thacker et al.\ 2000 for a careful discussion of this
point). Hereafter we denote the arithmetic average of the kernels as
$\overline{W}$.

The equation of motion is derived from the identity
\begin{equation}\label{ident}
{ \nabla P \over \rho}=\nabla {P \over \rho} + {P \over \rho^2} \nabla
\rho,
\end{equation}
where $P$ is the pressure.
Upon performing the standard SPH substitutions and making the following
replacement,
\begin{equation}
\nabla_i \overline{W}({\bf r}_i-{\bf r}_j,h_i,h_j) = -\nabla_j
\overline{W}({\bf r}_i-{\bf r}_j,h_j,h_i) + {\cal O}(\nabla h), 
\end{equation}
this becomes (neglecting  ${\cal O}(\nabla h)$ terms)
\[
{d {\bf v}_i \over dt}=\sum_{j=1}^N { {\bf f}_{ij} \over
m_i}=
- \sum_{j=1,r_{ij}<2h_i}^N
 m_j {P_i \over \rho_i^2} \nabla_i
\overline{W}({\bf r}_i-{\bf r}_j,h_i,h_j)
\]
\begin{equation} \label{2}
\;\;\;\;\;\;\;\;\;\;\;\;\;\;\;\;\;\;\;\;\;\;\;\; +
\sum_{j=1,r_{ij}<2h_j}^N
m_j {P_j \over \rho_j^2} \nabla_j
\overline{W}({\bf r}_i-{\bf r}_j,h_j,h_i),
\end{equation}
where ${\bf v}_i$ is the velocity of particle $i$ and ${\bf f}_{ij}$
is the force on particle $i$ due to particle $j$.

Finally, the energy equation used is that of Benz (1990),
\begin{equation}
{d \epsilon_i \over dt}=\sum_{j,r_{ij}<2h_i}^N m_j ({P_i \over \rho^2_i} +
{ \Pi_{ij} \over 2})
({\bf v}_i-{\bf v}_j) \cdot \nabla_i \overline{W}({\bf r}_i-{\bf r}_j,h_i,h_j),
\end{equation}
where $\epsilon_i$ is the internal energy of particle $i$ and
$\Pi_{ij}$ is the artificial viscosity between the two 
particles.  Note that while this equation is asymmetric in
the particle indices, it  still conserves energy to a high degree of 
accuracy. The single-sided approach is particularly helpful here since it
prevents spurious heating and cooling of collapsed, dense objects.
Radiative cooling is calculated for a constant $0.05 Z_\odot$ metallicity
gas, and the precise cooling table is interpolated from Sutherland \&
Dopita (1993). Because of the enormously fast change of cooling rates with
temperature we calculate the exact amount of cooling using an integral
formalism (see Thomas \& Couchman 1992) that assumes constant density
over a given time-step.

\subsection{Identification of Dwarf Galaxies}

To study the impact of galaxy outflows on nearby density
perturbations we must first know where and when
a galaxy is formed in our simulations.   As many of the objects of 
interest are at the limit of the resolution of the simulation,
complex criteria that attempt to capture in detail
the formation of massive stars in protogalaxies and the transfer of
momentum from stellar winds and supernovae into galaxy outflows
are not warranted.  Such criteria inevitably introduce a number of
parameters that affect the results in ways that are difficult to
disentangle and inevitably subject to fine tuning.
Thus in this study we adopt a model that strives to be 
as simple as possible and parallels our previous semi-analytical
work on this question (SB).

A common approach to object identification is to employ a
friends-of-friends group finding algorithm, which links together all
pairs of particles with separations less than some ``linking length''
$r_L$.  While this technique has been widely tested and known to give
good results (see eg.\ Davis et al.\ 1985; Lacy and Cole 1994) it is
much too computationally expensive to implement regularly during a
simulation.

Hence, we instead choose to identify collapsed protogalaxies by a
simple density cut.  Every ten time cycles we find the most dense gas
particle in each zone of the simulation that has not previously been
identified as belonging to a galaxy and make a list of all zones in
which these particles have overdensities above a threshold value,
$\delta_c$.  To account for galaxies centered near zone edges, we
exclude all zones with maximum densities lower than those of any of
their neighbors.  The points in the remaining zones are then taken to
be candidates for the centers of new galaxies.

Around each of these points we sort all the neighboring baryonic
particles according to their radial distance and continue to add 
particles (1, 2, 3.... $N$)
until $\delta_c \overline \rho_b \geq N m_g/ \frac{4 \pi}{3} 
r_N^3$, where $m_g$ is the mass of each gas particles,
and $r_N$ is the distance from particle $N$ to the center of the 
candidate galaxy.
We then compute the center of mass of this configuration,
resort the particles according to the distance from the center of 
mass and again add particles until  
$\delta_c \overline \rho_b \geq N m_g / 
\frac{4 \pi}{3} r_N^3$.  If the resulting object has 
a minimum mass of $2.5 \times 10^7 M_\odot$  corresponding
to 52 particles 
(the number of objects in the smoothing kernel of the SPH solver),
the object is considered to be a galaxy with baryonic mass $m_g N$,
otherwise the peak is below the resolution of the
simulation and is ignored.

While this prescription is both easy to implement computationally and
captures the essence of dwarf galaxy formation, it is important that
we understand how it compares with more standard methods of
object-identification. To make this comparison, we have carried out a
simple test simulation, with no star formation or outflows.  In this
simulation, as well as in the more detailed simulations described
below, we fix the cosmological parameters to correspond to the
observationally favored $\Lambda$CDM cosmology.  In this case the
current non-relativistic matter, vacuum, and baryonic densities in
units of the critical density are $\Omega_M = .35$, $\Omega_\Lambda =
.65$, and $\Omega_b = .06$, while the Hubble constant, amplitude of
mass fluctuations on the 8 Mpc/$h$ scale, and Cold Dark Matter ``Shape
Parameter'' are taken to be $H_0 = 100 h = 65$ km/s/Mpc, $\sigma_8 =
0.87,$ and $\Gamma = 0.18.$

\begin{figure}
\centerline{\psfig{file=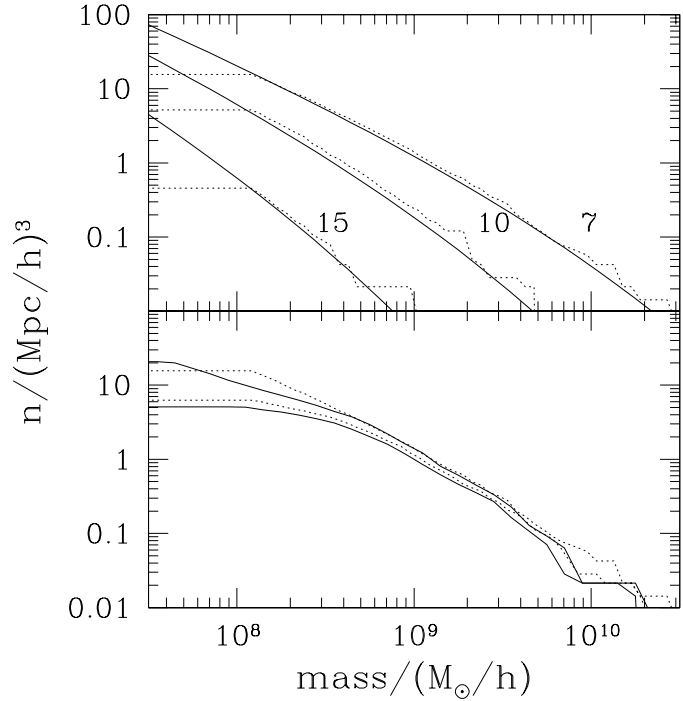,width=3.8in}}
\caption{Identification of Galaxies.  {\em Top}: The number density
of dark matter halos predicted by Press-Schechter theory (solid lines) 
and the number density identified by the friends-of-friends group finding 
algorithm (dotted lines) at three different 
redshifts.  {\em Bottom}:  The upper and lower dotted lines show 
the friends-of-friends number density of dark-matter halos 
with a linking length of $500^{-1/3}$ of the inter-particle spacing 
and baryonic clouds 
with a linking length of $4000^{-1/3}$ of the inter-particle spacing 
respectively. The solid
curves show the number density of galaxies as identified by a
density cutoff of 200 (upper) and 500 (lower).}
\label{fig:tagging}
\end{figure}

In the top panel of Figure \ref{fig:tagging}, we compare the number 
density of collapsed dark matter halos as identified by a friends-of-friends
group-finding algorithm with a linking length of $500^{-1/3}$ of the
average inter-particle spacing to the number density of 
halos as predicted analytically from a Press-Schechter approach.
In the numerical case,  the mass has been shifted
by a factor of $\Omega_M/(\Omega_M-\Omega_b)$ to account for the
fact that the analytical expression accounts for the total 
dark matter plus baryonic mass of the
object.  Here we see that while there is some discrepancy at the 
largest mass-scales due to small-number statistics, this method of 
identification does well at all redshifts.

In the lower panel, we compare the number of dark matter halos 
with galaxies as identified from the distribution of the gas.
Again the masses have been shifted by a factor
of $\Omega_M/(\Omega_M-\Omega_b)$ in the dark matter case and
by $\Omega_M/\Omega_b$ in the case of the gas.  
Because of the increased condensation of gas
relative to dark matter due to radiative cooling, a friends-of-friends
linking length that is half that of the dark matter is necessary to give 
good agreement for all but the smallest mass scales.
In this figure we also show the number density of galaxies as identified
by the density-cut method described above, with two threshold
values of $\delta_c = 200$ and $\delta_c = 500.$  Again, both
these values give reasonable agreement with the friends-of-friends
dark matter and friends-of-friends gas particle identification schemes
over a large range of mass scales.  The identification of objects
is quite dependent on this threshold on the lowest mass scales however,
with the $\delta_c = 200$ case slightly over-predicting the number of
objects relative to the dark-matter friends-of-friends case,
while the $\delta_c = 500$ cut gives slightly fewer low-mass objects
than a friends-of-friends identification in the gas.
In order to be conservative in the number of outflows 
included in our simulations, as well as reproduce as closely as possible
the results of a friends-of-friends identification of regions of
virialized {\rm gas}, we therefore adopt the higher density cut
in our main simulations.

\subsection{Formation and Propagation of Galaxy Outflows}

Having developed a means of quickly identifying newly-formed dwarf
galaxies, we must also implement a model for star formation and
outflows in these objects.  For our model of star formation, we assume
a simplified picture in which a fraction $\epsilon_{\rm sf}$ of the
gas in a collapsed object is converted into stars in a single, initial
burst of star formation. As all subsequent star bursts within a galaxy
are likely to be much smaller and less efficient at generating
outflows, this can be thought of as a conservative lower limit on the
number of outflows formed.

In order to avoid multiple identifications of the
same galaxy at each time step, and hence excessive
star formation, we tag all particles that have previously 
been identified as part of a galaxy.  Star formation is only
implemented in cases in which over one third of the galaxy 
and at least 52 particles are untagged material that has not 
been previously considered as part of a collapsed object.
Finally, in order to avoid excessively large outflows, we impose 
an upper limit of 1000 gas particles, corresponding to $5 \times 10^8 M_\odot$ 
of gas that can be converted into stars in any one star-burst.

As we are interested only in the number and spatial
distribution of galaxies, and are not attempting to reproduce
their properties in detail, we simply take these stars to be formed 
out of the inner $\epsilon_{\rm sf}$ fraction of the gas particles,
located closest to the center-of-mass.  Apart from being easy to implement, 
this prescription also helps to delay excessive slowing-down of the
simulation as it reduces the net number of SPH neighbors in the densest regions.

Construction of outflows from these starbursting galaxies is likewise
carried out in a simple manner, although in this case a slightly more
sophisticated approach is necessary in order to properly reproduce
outflows with realistic properties.  The structure of outflows has
been studied in detail by Mac Low \& Ferrara (1999) and Ferrara \&
Tolstoy (1999), who conclude that efficient ejection of the
interstellar medium or ``blow-away'' in disk galaxies occurs only 
in objects associated with halos with masses
$\lesssim 10^7 M_\odot$).  
In larger disks within halos in the mass range $10^7
M_\odot \lesssim M \lesssim 10^9 M_\odot$, a ``blow-out'' occurs in
which the super-bubbles around groups of Type II supernovae punch out
of the galaxy, expanding freely perpendicular to the disk and
shocking the surrounding IGM while failing to excavate
the interstellar medium of the galaxy as a whole (see also Martel 
\& Shapiro 2000).  This scenario goes
far in reconciling observations of expanding shells around dwarf
galaxies (see eg.\ Axon \& Taylor 1978; Marlowe et al.\ 1995; Heckman
1997; Martin 1998) and spectroscopy of high-$z$ galaxies (eq.\ Frye \& 
Broadhurst 1998; Pettini et al.\ 2001 ) with observations of multiple
episodes of star formation in dwarf spheroidal galaxies (Smecker-Hane
et al.\ 1994; Grebel 1998) some of which even suggest that many of
these objects are gas-rich, but with extended HI envelopes (Blitz \&
Robishaw 2000).

Guided by this blow-out scenario, we model
the sub-resolution physics of outflow generation by rearranging
the IGM surrounding an object into a galactic wind while leaving
the central object intact.
We construct our outflows from all
gas particles with $r_N < r < r_O,$ where $r_O$ is the maximum of
twice the radius of the object identified as a galaxy and the radius
within which 104 additional particles are located ($r_O = {\rm max}
\left[2 r_N, r_{N+104} \right]$).  All the objects within the galaxy radius,
$r_N$, on
the other hand, are left unchanged apart from the simple conversion
of a fixed fraction of gas particles to stars.  These outflow
particles are then arranged on two concentric spherical shells of
radius $r_O$ and $0.9 r_O$.  This multi-shell structure assures that
the outflows will be sufficiently well-resolved radially to be
reasonably treated by the SPH solver.  Sufficient resolution along
each of the shells is assured by arranging the particles in each of
the shells to be anti-correlated such that no two particles are
within a distance of less than one half of the average spacing
between neighbors.  This minimizes the sub-random particle distribution
that arises naturally within SPH. The particles in both shells are then 
given velocities 
\be 
{\bf v}_{i,{\rm shell}} = {\bf v}_{\rm cm} + 
{\bf \hat r_i} {v}_{\rm rad} + {\bf v_i}_{\rm ang}, 
\ee 
where ${\bf v}_{\rm cm}$ 
is the center of mass velocity of the galaxy, $v_{\rm rad}$ is a
boost in the radial direction due to the outflow, and ${\bf v_i}_{\rm
ang}$ is an axial velocity necessary to conserve angular momentum.

Just as we have assumed in all cases that a fixed fraction of the
the gas in the object, $\epsilon_{\rm sf}$,  forms stars, we now assume 
that a fixed fraction  of the energy from the resulting supernovae,
$\epsilon_{\rm wind},$  is channeled into the outflow. 
Following the prescriptions adopted in SB we estimate
that one supernova occurs for every 100 $M_\odot$ that form 
stars (see eg.\ Gibson 1997) and that each of the
supernovae has a kinetic energy of output of $2 \times 10^{51}$ 
ergs, to take into account the contribution from stellar winds.
The radial velocity of the shell is then taken such that
$1/2 N_{\rm shell} m_g v_{\rm rad}^2= {\rm KE} - {\rm PE}$ where 
$N_{\rm shell}$ is the number of particles used to construct the outflow,
KE is the kinetic energy channeled into the outflows,
and PE is the potential energy to move the particles from
their original locations to the two concentric shells.
Approximating the total mass of the galaxy as
$N m_g \Omega_M/\Omega_b$, this gives
\ba
v_{\rm rad} & = & 200 \, {\rm km/s} \, \left( \frac{N}{N_{\rm shell}} \right)^{1/2}
\left[
\frac{\epsilon_{\rm sf}}{0.1} \frac{ \epsilon_{\rm wind} }{0.1} 
 - 2.1 \times 10^{-10} \right. 
\nonumber \\ 
  &  & \, \, \, \times  \left.  \frac{m_g}{M_\odot}
\frac{\Omega_M}{\Omega_b} \left( \sum_{i=N+1}^{N+N_{\rm shell}} 
\frac{1 {\rm kpc}}{r_i} - \frac{1 {\rm kpc}}{r_O} \right) \right]^{1/2}. 
\ea
Note that this method relies on an un-softened potential- energy
estimate so that the shell does not have an excess of kinetic
energy as would result from a softened potential-energy approach.
Finally, to compute ${\bf v_i}_{\rm ang}$ we take the angular momentum
vector of the particles used to construct the outflow about the center
of mass, and apply a solid body rotation to the shell in the same
direction and with the same magnitude:
\be
{\bf v_i}_{\rm ang} = \frac{{\bf L} \times {\bf r_i}_{\rm shell}}
{(r_O)^2 N_{\rm shell}},
\ee
where ${\bf L} = \sum_{i = N+1}^{N+N_{\rm shell}}
	{\bf r_i} \times {\bf v_i}.$

In order to compare this outflow scheme with analytical estimates, we
have carried out a test simulation in which we fix $\epsilon_{\rm sf}
= \epsilon_{\rm wind} = 0.1$ and study the collapse and outflow
evolution of a spherical overdensity.  In a 1.8 Mpc cubic comoving
region containing $2 \times 50^3$ particles, we arranged the matter
within a 300 kpc radius into a spherical ``top hat'' overdensity, such
that it would collapse virialize at a redshift of two.  For simplicity
we considered an Einstein-de Sitter cosmology in which $\Omega_M =
1$, $\Omega_\Lambda = 0$, $\Omega_b = 0.06,$ and $h = 0.5.$ With
these parameters, the top hat contained approximately 2469 gas and
dark matter particles of masses $2.1 \times 10^6 M_\odot$ and $1.3 \times 10^5
M_\odot$ respectively, resulting in a total mass of $8.5 \times 10^9
M_\odot.$  While the original baryonic mass of this
perturbation was $4.5 \times 10^8 M_\odot$, only approximately half of
this mass collapsed into the final galaxy, due to the reduction in
accreted gas caused by the expanding shell.
This effect was also observed in our larger simulations and is
discussed in more detail below.

\begin{figure}
\centerline{\psfig{file=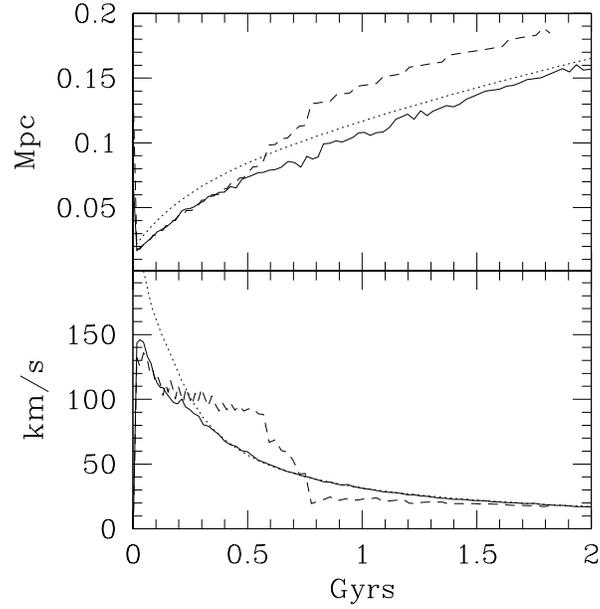,width=3.8in}}
\caption{Comparison between SPH simulations and analytical modeling.
The solid lines and dashed lines correspond to outflows from 
the top-hat within the Hubble flow and pure top-hat simulations
respectively, 
as discussed in \S2.3; while the dotted lines correspond to the thin-shell 
model adopted in SB.  In all cases the overdense region is a 300 comoving
kpc sphere that collapses at $z=2$ in an Einstein de-Sitter cosmology.
The time derivative of the radius in the pure top-hat does not correspond
to the velocity as the peak bin shifts in this simulation as the shock
passes through the empty region.}
\label{fig:oneflow}
\end{figure}

The trajectory of the outflow is shown as the dashed lines in
Figure \ref{fig:oneflow}.  At each time we calculated the velocity and
radius in the radial bin that contained the highest momentum.  The
sharp change in the radius at 0.5 Gyrs is due to the presence of an
extremely low-density gap left behind after the top-hat collapses.
When the shock moves into this region, the low density gas is heated
and accelerated to velocities greater than that of the dense material,
while conserving overall radial momentum.  As a result the material
spreads out and the peak becomes less defined, shifting from the
center to near the front of the shock.  This jump is particularly
clear when contrasted with the dotted lines, which show
the analytical solution taken from SB
for a model with $\epsilon_{\rm sf} = \epsilon_{\rm wind} = 0.1$
and an object with a total mass of $8.5 \times 10^9 M_\odot$.  
Here the shock is modeled as a thin spherical
shell of material which is accelerated due to internal pressure and
escapes from the gravitational pull of the halo into the Hubble Flow.
Note that although the SB model also takes into account the Compton drag
due to the scattering of electrons against CMB photons, this
contribution is negligible at $z=2.$

While the high velocities in the analytical solution
at early times ($\leq .2$ Gyrs) are due
to the pressure driven nature of the early outflow solution, the 
discrepancies at later times are clearly a relic of the existence of
the unnatural empty region in the SPH simulation.  
In order to be able to more directly compare
our simulations to the thin shell model then, we conducted a final
test in which we excised the central collapsed object and early outflow
from this empty region and superimposed it on a uniform 1.8 Mpc cubic
box of particles expanding with the Hubble flow.
The results of this simulation are shown as the solid lines,
for which the velocities agree almost exactly for all times
after the early pressure-driven phase.  
Note also that 
that the early shell velocities in all of these models are consistent with
those observed in Lyman break galaxies at similar optical radii
(Pettini et al.\ 2001).

\section{Results and Discussion}

In order to evaluate the impact of galactic outflows on structure
formation, we have conducted two simulations: a main outflow run with
$\epsilon_{\rm sf} = 0.1$ and $\epsilon_{\rm wind} = 0.1$
corresponding to the fiducial model examined in SB, and a comparison
``no-outflows case'' in which star formation is implemented with the
same efficiency, but no outflow particle rearrangements and velocity
boosts are implemented.  Our philosophy is not to attempt to refine
these parameters in order to best reproduce the observed properties of
dwarf galaxies, but rather to study the qualitative features that
arise in a simple, conservative model of galaxy outflows.  Once the
SPH particles in a given region reach their minimum allowed smoothing
length, further clustering causes the number of SPH neighbor particles
to rise without limit. Although the transfer of gas to stars helps to
alleviate the subsequent algorithmic slow-down, the $N^2$ nature of
clustered regions leads to a significant load imbalance in the
parallel code. Hence we limited the evolution of the simulations to
$z=6$, which is more than sufficient to detail the properties of the
dwarf galaxy distribution under the influence of outflows. Evolution
to this epoch required over 2000 steps, with a minimum time-step of
0.4 Myr. Notably the run with outflows required 5 per cent fewer
time-steps because dense regions, which usually dictate the shortest
time-steps, form with less efficiency.  Each simulation required
approximately ten days of computing time on a 16 node Sun E6500
server.

\subsection{Global Properties and Observational Checks}

\begin{figure}
\centerline{\psfig{file=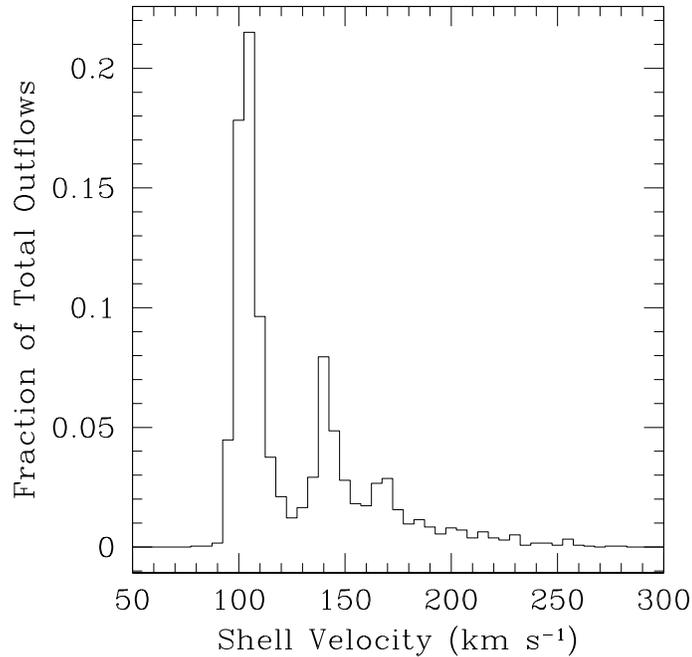,width=3.8in}}
\caption{Histogram of outflows as a function of initial shell velocity.  
The lower 100 km/s peak is a relic of the minimum galaxy mass in the
simulation.  There are 2367 outflows in the (5.2 Mpc/$h$)$^3$ comoving 
volume by a redshift of 6. }
\label{fig:of}
\end{figure}

In Figure \ref{fig:of} we plot the distribution of outflows 
as a function of initial velocity.  While the lower
peak in this plot is almost certainly a relic of the minimum
mass scale at which we identify objects as galaxies,
a large fraction of the winds are generated with significantly
higher initial velocities.  All velocities are consistent with 
observations of high-redshift dwarf galaxies (Frye \& Broadhurst
1998; Warren et al.\ 1998; Pettini et al.\ 2001). 

\begin{figure}
\centerline{\psfig{file=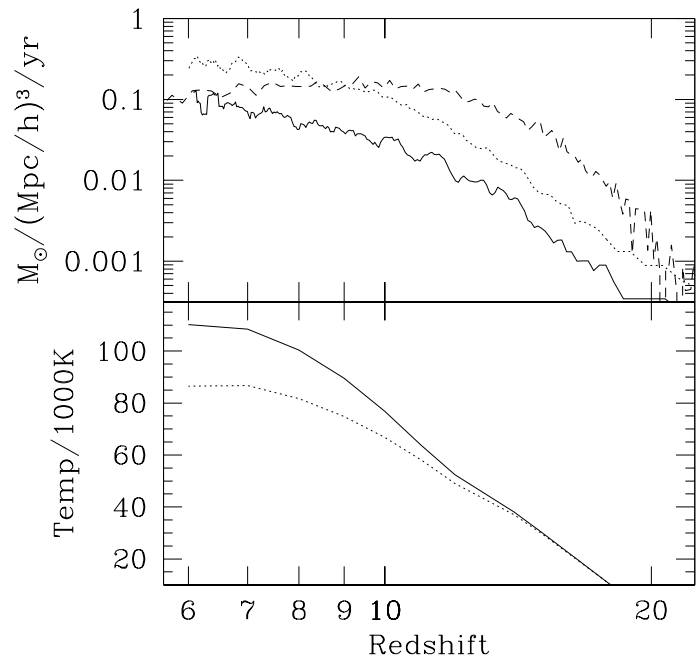,width=3.8in}}
\caption{{\em Top:} Global star formation rate.
The star formation rate in the outflows simulation is given by the
solid line while the star formation only simulation is represented by the
dotted line.  For comparison, the dashed line shows the semi-analytical
star formation rate calculated in SB for the fiducial outflows model.
{\em Bottom:} Mean temperature of the gas 
in the outflow (solid) and star formation only (dotted) simulations.}
\label{fig:sfr}
\end{figure}

In the top panel of
Figure \ref{fig:sfr} we show the overall star formation rates (SFR) 
in these simulations. While these rates are equivalent at extremely 
early times, almost immediately the 
addition of outflows to the simulation greatly reduces the mass of gas 
passing the density-cut criteria, reducing the SFR in the outflow simulation
by a factor of approximately
three. For comparison, we also show on this plot the SFR as calculated
in the semi-analytical model described in SB.  Here star formation
occurs at slightly higher redshifts, due to the stringent density-cut
criteria chosen to identify galaxies in our simulations.  At late
times, however, both these rates become roughly similar.
These values are also roughly consistent with measurements
at redshifts 3 and 4 and extrapolations to the redshift range
in our simulation (see eg.\ Steidel et al.\ 1999;  Percival, Miller, 
\& Ballinger 1999; Huges et al.\ 1998; Gallego 1995)

Finally, in the lower panel of this figure, we show the 
average temperature of the gas particles in the simulation.
Here we see that outflows have only a secondary
impact, the majority of gas heating
being the result of energy released by gravitational collapse 
and virialization.  Nevertheless, the overall increase in IGM 
temperature pushes our model in a direction that is
favorable to X-ray background considerations (Pen 1998),
and the overall degree of heating is well within other
observational limits.  A main probe of this heating is the
degree of spectral distortions in the Cosmic Microwave 
Background, due to scattering of microwave photons off the hot
ionized gas.  The magnitude of these distortions is given by the 
Compton-$y$ parameter which is the convolution of the optical 
depth with the electron temperature along the line of sight 
(see eg.\ Zel'dovich \& Sunyaev 1969; Sunyaev \& Zel'dovich 1972).  
In our models the contribution to this effect from
redshifts above 6 is $y= 1.3\times 10^{-6}$ in the outflows case
and $y= 1.0\times 10^{-6}$ in the star formation only case.  
These values are much less than the observational constraint of 
$y \leq 1.5 \times 10^{-5}$ (Fixsen et al.\ 1996). 

\subsection{Suppression of Dwarf Galaxy Formation and Baryonic Stripping}

\begin{figure}
\centerline{\psfig{file=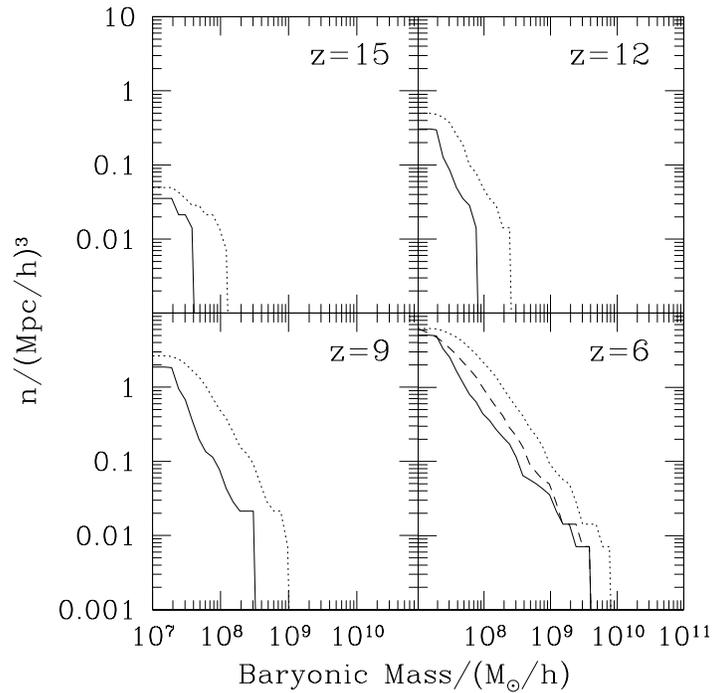,width=3.8in}}
\caption{Number density of galaxies, now in terms of baryonic mass.
In each panel the solid lines are from the simulation with galaxy outflows,
and  the dotted lines are from the star formation only simulation.
Notice that there is both an overall offset in mass between
the two runs, as well as an additional suppression of halos
on the $\sim 10^8 M_\odot \approx 10^9 M_\odot
\Omega_b/\Omega_0$ scale.  For comparison,
the dashed line at $z=6$ shows the no-outflows run shifted in mass
by a factor of two.}
\label{fig:number}
\end{figure}

The great reduction in star formation in the outflows case
is indicative of a large difference in the total mass in galaxies
between the two runs.   In order to understand how this difference
is distributed over the range of galaxy masses, 
in Figure \ref{fig:number} we compare the number-densities of
galaxies in both runs at a number of redshifts. 
Again, we use the density-cut method of identifying 
objects with a threshold value of $\delta_c = 500,$ 
and shift the total gas mass by a factor of $\Omega_M/\Omega_b$ to
approximate the total mass of the object.

Here we see that outflows have two major effects on the distribution
of objects.  First, as in the test case discussed in \S 2.3,
they somewhat restrict the amount of gas that each
object can accrete, shifting the number distribution to 
smaller masses at all mass scales.  Thus while galaxies in the $\sim
10^7 M_\odot$ to $\sim 10^{10} M_\odot$ range may have been unable to
eject large fractions of their own interstellar medium, the strong
winds of IGM material they are able to ``blow-out'' at early times
greatly reduce the amount of gas they accrete over their lifetimes.
This is similar to the mechanism described by Dekel \& Silk (1986) in
which high dark matter fractions are caused by supernova-driven winds
from high-mass stars formed early in the history of a dwarf galaxy,
resulting in a systematic bias towards higher mass-to-light ratios.

In addition to this mass shift, Figure \ref{fig:number} shows a large
deficit in the overall number of objects with masses below
$5 \times 10^9 M_\odot$.  At $z=6$ for example, even when
accounting for a $\sim 50 \%$ shift in masses between the two runs,
there are less than half the number of galaxies in this 
range in the outflows simulation than in the no-outflows case.  This
difference is extremely suggestive of the ``baryonic stripping''
mechanism identified in our previous studies, in which outflows from
early galaxies remove the gas from nearby overdense regions that would
have otherwise later formed into dwarf galaxies.  In these regions,
the gas is stripped away from the collapsing dark matter and ejected
into space along with the outflow, leaving behind only an empty ``dark
halo'' of non-baryonic matter.  In SB, baryonic stripping
reduced the number of $\sim 10^9 M_\odot$ objects by $\sim 75\%.$

\begin{figure}
\centerline{\psfig{file=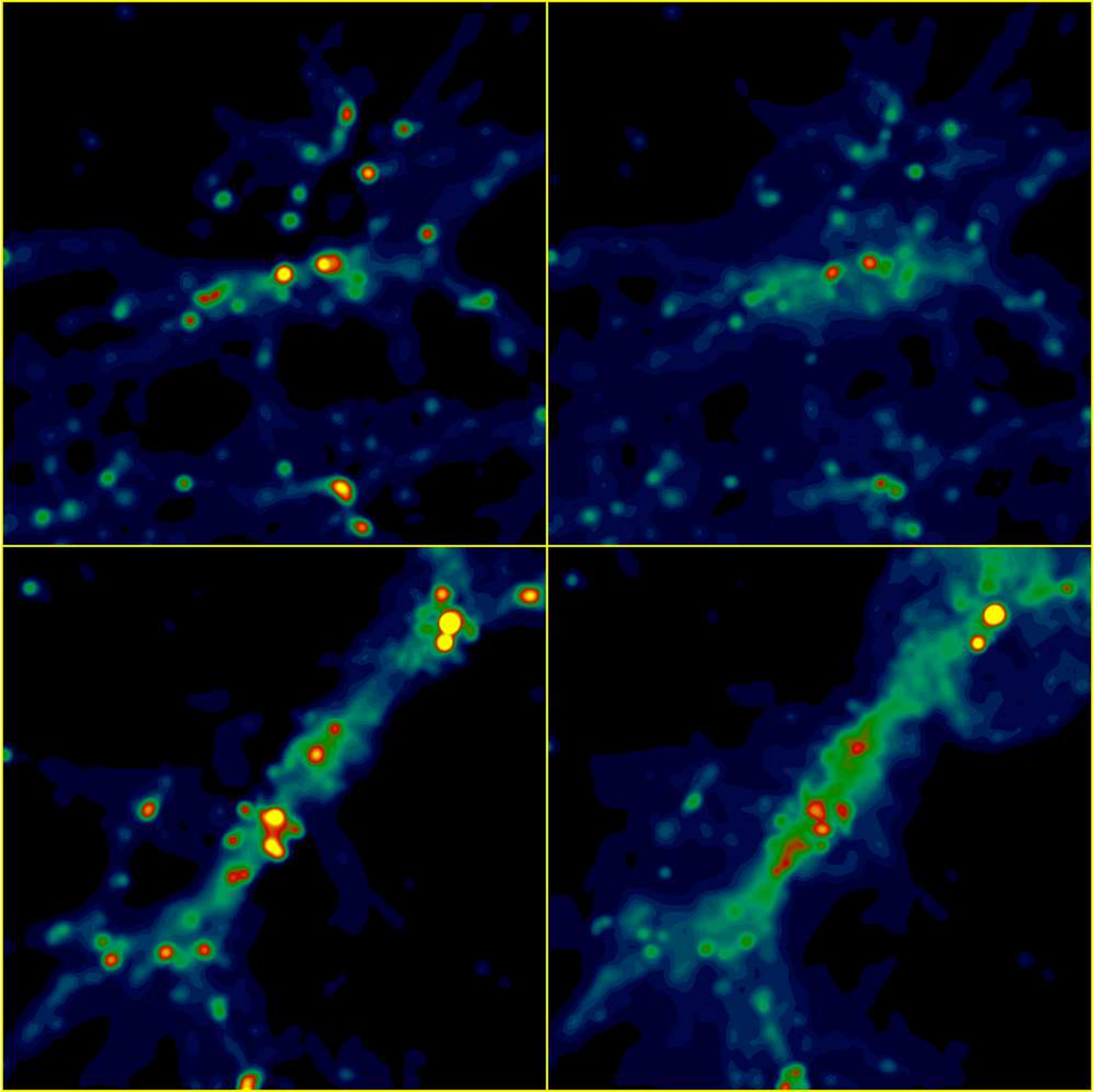,width=6.8in}}
\caption{Mass contours of two 80$^3$ kpc$^3$ physical (560$^3$ kpc$^3$
comoving) regions from each of the simulations at a redshift of 6. 
The colors are arranged in equal contours from $1.3 \times 10^{10}  M_\odot$ to$6.5\times 10^{11} M_\odot/({\rm comoving} \, {\rm Mpc})^2$,
while the mean density in this cosmology is  
$9.3\times 10^{9} M_\odot/({\rm comoving} \, {\rm Mpc})^2$.
The upper two panels show a typical region of 
galaxy formation.  In the lower two panels, we compare
two regions of vigorous galaxy formation.
In each pair, the left panel is taken from the no-outflows run,
while right panel is from the run with galaxy outflows.}
\label{fig:color}
\end{figure}

In order to study this mechanism further in our SPH simulations, we
have selected two 60$^3$ kpc$^3$ physical (420$^3$ kpc$^3$ comoving)
regions at a redshift of 6: one a typical region of galaxy formation,
and the second a more extreme example with higher-velocity outflows.
These regions have been extracted from both simulations and used to
construct contour plots of total gas mass projected in the
$z$-direction, shown in Figure \ref{fig:color}.

In this figure we can see both the reduction of gas accretion in
outflowing dwarfs, and the baryonic stripping of neighboring objects.
The typical galaxy forming region shown in the upper two panels
emphasizes the first of these two mechanisms.  While both runs contain 
the same number of dense peaks in the gas distribution, these are somewhat 
larger and more concentrated in the no-outflows case.  Thus the two largest
peaks in the center of the volume have much the same positions and central 
mass profiles in both cases, while much of the mass on the outer edges
has been ejected into a large cloud rather than accreted
by the galaxies themselves.  Note also that no outflows 
occurred in the less dense regions such as that near the bottom of 
these panels,  and thus the distribution of matter in this
area is completely equivalent between the two runs.

In the extreme outflow regions shown in the bottom two panels
of Figure \ref{fig:color}, reduced accretion is again visible,
but now complemented by several examples of baryonic stripping.
Notice, for example, that the two smaller-mass peaks to the left of
the central object in these panels have been almost completely disrupted 
by the spherical wind emanating from this object.  Notice also that 
this mechanism is extremely sensitive to the relative formation times 
between objects.  Thus, while the two peaks directly above and to the 
right of the central object are of similar sizes, the outflow
from the first of these objects to form has completely suppressed
the other's formation.  A careful inspection of these panels
uncovers several other examples of baryonic stripping and slowing of 
accretion, although again, the lowest-density regions are completely
equivalent.

In order to further quantify this suppression of objects, we make
use of the fact that while outflows greatly affect the number
and distribution of galaxies, their effect on the sizes and positions
of collapsed dark matter halos is only of secondary importance.
Using the friends-of-friends group finder method with a linking 
length of $500^{-1/3}$ of the average inter-particle spacing, we have
computed lists of dark matter halos as described in \S 2.2.
For each of these objects we identify a center-of-mass
position along with an average distance from the center of mass to
the particles in the halo, $R_A.$  We then calculate the
mass of galaxies lying within each halo by tagging all ``galaxies,''
as defined by our density cut criteria,
whose center of mass lies within $f R_A$, where $f$ is some
fraction.  Here we fix $f = 2.0,$ although varying this quantity
from 1.5 to 2.5 has little effect on our results.  

\begin{figure}
\centerline{\psfig{file=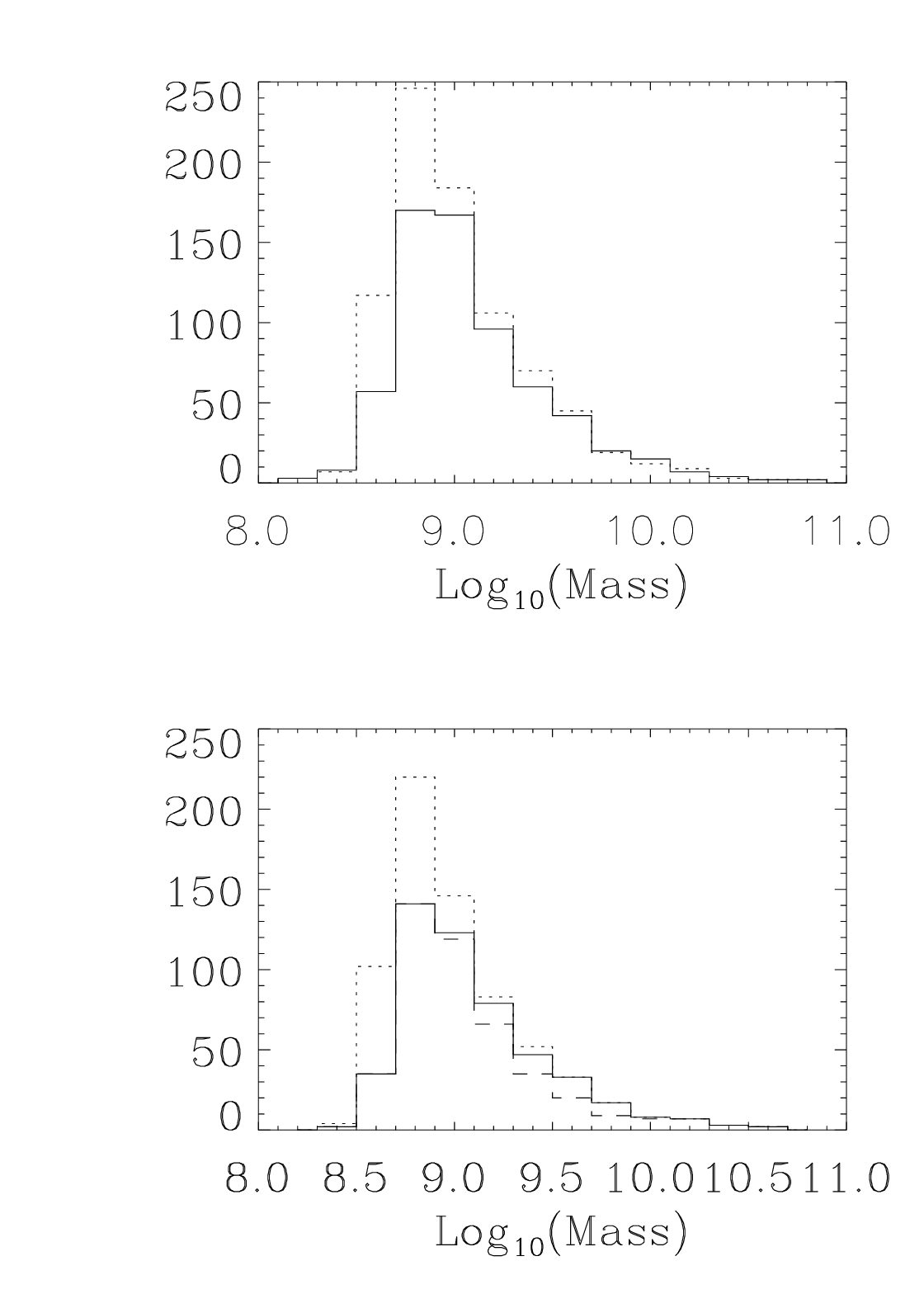,width=3.8in}}
\caption{Baryonic stripping and empty halos.  {\em Top:}
Histograms of dark matter halos that contain at least one
condensed galaxy.  The solid lines correspond to the outflows run
and the dotted lines to the star formation only case.
{\em Bottom:}  Histograms of dark halos that have been correlated
between the two runs.  The dotted line shows all correlated
halos that contain a galaxy in the star formation only run and
the solid line shows all halos with galaxies in both simulations.
Finally, the dashed line shows all halos with galaxies in
the star formation only run, and galaxies with at least $1/4$
of that mass in the outflows run.}
\label{fig:empty}
\end{figure}

In the upper panel of Figure \ref{fig:empty} we plot
the number of halos as a function of halo mass
that contain at least one galaxy as defined in \S2.2.  
Here we see that the no-outflows run has
a significantly larger population of halos  in the $\sim 10^9 M_\odot$ 
range that contain galaxies. Note that this deficit cannot be 
caused simply by a reduction of accretion, as this histogram
contains all halos that host a galaxy of any size.

This difference can also not be attributed simply to a difference in the
number of dark matter halos in both simulations.  In
the lower panel of Figure \ref{fig:empty} we have correlated
the halos between the simulations, requiring that their
center-of-masses be located within 1 kpc of each other.  
In this case we show the number of correlated halos with
a galaxy in the outflows simulation, compared the 
correlated halos with galaxies in both simulations. 
While the overall number of halos is reduced slightly due to
small differences in the halo distribution, we see again that
a large number of the filled halos in the no-outflows case 
are empty in the corresponding outflows run.
Finally, we compute the number of halos that contain
a galaxy in the no-outflows run, but contain less than a quarter of
that mass in galaxies in the outflow case.  These examples of 
severe suppression persist up to the $10^{10} M_\odot$ scale.

\begin{figure}
\centerline{\psfig{file=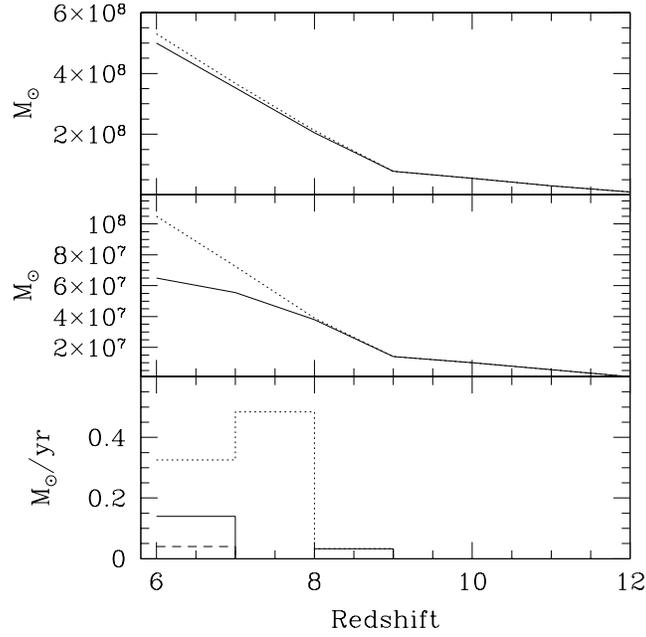,width=3.5in}}
\caption{Baryonic stripping in a typical halo.
The upper and center panels show the total dark matter and baryonic
mass respectively, within a 10 kpc/h comoving sphere centered around a
halo that experiences baryonic stripping.  The solid lines correspond
to the fiducial outflows run, while the dotted lines are taken from
the run with no outflows.  In the lower panel the solid and dotted
lines show the star formation rate in the 6.5 physical kpc/h sphere
representing the environment of the halo in the no-outflow and
fiducial runs respectively, while the dashed line shows the
star-formation rate within the 10 kpc/h object itself in the
no-outflows case.  The are no stars formed within the object itself in
the fiducial run.}
\label{fig:onemore}
\end{figure}

Finally, we chose a typical halo that is empty in the fiducial case
and filled in the no-outflows run and study its history in greater
detail.  In Figure \ref{fig:onemore} we plot the total dark and gas
matter masses within the object itself, defined as the mass within a
10 comoving kpc $h^{-1}$ sphere centered around the center of mass of
a dark matter halo with a mass of $5 \times 10^8$ a redshift of 6.
When then compare the evolution of this mass to the star formation
rate within a fixed 6.5 kpc $h^{-1}$ sphere representing the object's
immediate environment.  At early times, the evolution of the object is
the same in both runs, until the initial burst of nearby
star-formation at a redshift of 9.  As the outflow corresponding to
this burst sweeps though the halo, the total number of baryons
diverges between the two runs, while the number of dark matter
particles remains almost the same.  By a redshift of 6, the
no-outflows object has accreted almost twice the gas of the object in
the outflow run.  Note that this discrepancy is of a different nature
than the mass shift described in \S2.3 as it is caused by
environmental effects rather than winds emanating from the object
itself.  Indeed, no star formation occurs at any redshift $\geq 6$ in
the object in the fiducial run, and only the no-outflows object passes
the density cut to be considered a galaxy and forms stars.  Notice
also that while environmental effects are able to suppress this object
only in the outflows case, the total degree of environmental
star-formation is actually much greater in the no-outflows simulation,
as neighboring objects accrete gas more easily in this model.

\subsection{Theoretical Uncertainties}

While the simple ``blow-out'' scenario we have studied strives to
illustrate the qualitative effects of dwarf outflows on structure
formation, without attempting to reproduce the process in detail, it
nevertheless contains a number of free parameters that deserve some
scrutiny.  The most important and uncertain of these values for our
simulations is the product of $\epsilon_{\rm sf}$ and $\epsilon_{\rm
wind}$ which compounds the uncertainty in starburst size with the
uncertainty in the fraction of that energy that is channeled into the
winds. The star formation efficiency is one of the most complicated
and notoriously uncertain parameters in all of astrophysics, and as
such there is little we can do to definitively fix this number.  Our
value is thus simply chosen to be somewhat middle of the road among
estimates (eg. Tegmark, Silk, \& Evrard 1993; Majumdar, Nath, \& Chiba
2000) and consistent with observed SF rates at high redshift.

Similarly $\epsilon_{\rm wind}$ is largely uncertain,
and is chosen to be a somewhat conservative value, consistent with
previous estimates (eg. Heckman, Armus, \& Miley 1990; Scannapieco \&
Broadhurst 2000; Madau, Ferrara, \& Rees 2000).  Note that our choice
of $\epsilon_{\rm sf} \epsilon_{\rm wind} = 0.01$ results in outflow
velocities that are consistent with and even smaller than those
observed and does not cause excessive heating of the IGM beyond
Compton constraints, and thus we can have some confidence that this
is a reasonable value to consider.  Our goal, after all, is not to
present a definitive model of galaxy outflows, but rather an
illustrative example of the role that they play at high redshift: a
picture that will undoubtedly be refined as our theoretical and
observational understanding of galaxy outflows progresses.

While these uncertainties are unavoidable in any study of
high-redshift outflows, our model also contains an additional
parameter that is specific to how burst-mode
star formation is implemented. Since accretion occurs over
a number of time-steps it is difficult to ascertain
when an object has accumulated enough material to
be considered a `new galaxy', and thus to undergo a starburst. In our
fiducial and no-outflow models we implemented a ``one-third'' rule,
adding a starburst whenever a fraction $f_B = 1/3$
or 52 particles in a galaxy
were new material. The 1/3 value was initially chosen as it corresponds 
to a similar threshold used in the semi-analytic model developed in SB.
Setting $f_B = 0$ would result in each
galaxy undergoing a new starburst with each small amount of gas
accreted, constantly creating new stars and rearranging the
surrounding IGM material into outflows.  Adopting
a very high value would likewise result in an equally
unphysical situation in which galaxies were forced to remain quiescent
for a long latency period, accreting huge amounts of gas before they
finally explode in massive outflows.

In order to be certain that our simulations lie somewhere between these
unphysical extremes, we have carried out a number of tests runs in
a periodic comoving box of size $3.0$ $h^{-1}$ Mpc with
$2\times 96^3$ particles.  The mass resolution in this case is $3.9
\times 10^6 M_\odot$ for the dark matter and $7.8 \times 10^5 M_\odot$
for the gas.  We again fixed $\epsilon_{\rm sf} =
0.1$ and made four different outflow runs with $\epsilon_{\rm wind} =
0.1$, one with no lower limit on the fraction of new particles necessary
to be considered a new galaxy, and three runs where we allowed 
$f_B$  to vary between 1/4, 1/3, and 1/2.  Finally, we conducted a
comparison no-outflows test run with $\epsilon_{\rm wind} = 0$.

\begin{figure}
\centerline{\psfig{file=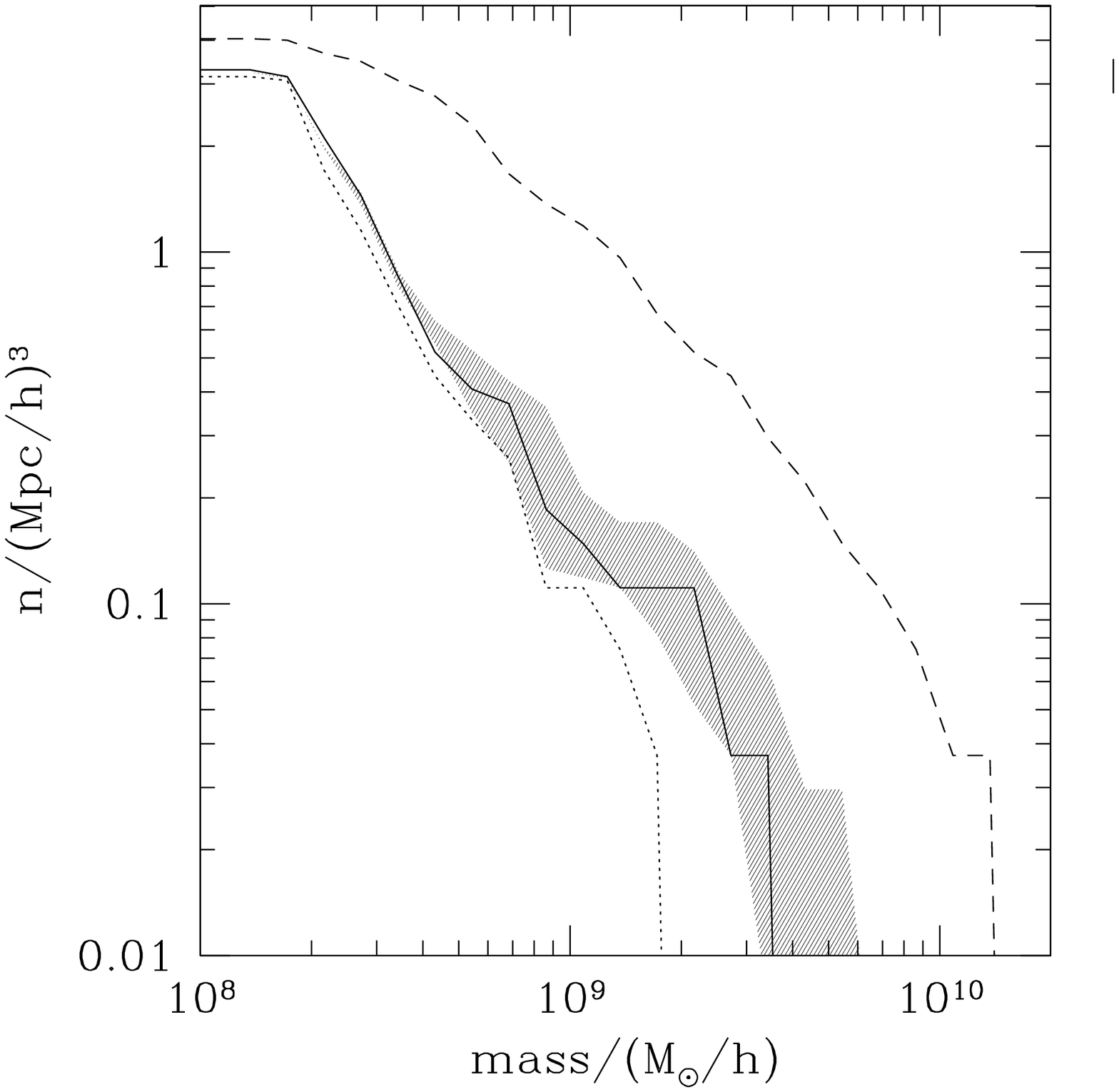,width=3.5in}}
\caption{Number density of galaxies in the various comparison runs
at a redshift of 6.  The solid line is from the outflow simulation with 
$f_B = 1/3$ and the shaded region is bounded by the results of varying
$f_B$ from 1/4 to 1/2.  The dotted line represents an outflow
simulation in which $f_B = 0$
while the dashed line shows the results of the test
run with no outflows.  Although varying the 
threshold value introduces some uncertainty in the number of
objects in the outflow runs, these changes are small when compared
with the no-outflows case.}
\label{fig:fraction}
\end{figure}

The resulting number distributions appear in Figure
\ref{fig:fraction}.  Here we see that while changing $f_B$ does
introduce some uncertainty in the number of objects formed in the
outflow simulations, these uncertainties are small in comparison with
the difference between these runs and the no-outflows case.  Furthermore
at all masses, the number of objects increases monotonically with
increasing threshold, thus demonstrating that even at a somewhat high
value of $f_B = 1/2$, unphysically powerful superwinds do not conspire
to increase or bias the impact of galaxy outflows from the behavior
seen in the less-punctuated models.  Relaxing the threshold entirely,
on the other hand, results in a huge reduction in the amount of gas
accreted by the galaxies, as outflows and IGM rearrangements are
implemented every few times steps.  A simple visual inspection of this
run confirmed that the majority of the galaxies in the volume were
little more than groupings of star particles, surrounded by concentric
spherical shells.  Based on these results, we can have some confidence
that by adopting an intermediate threshold value of 1/3, our simple
model is able to reproduce the qualitative features of the impact of
galaxy outflows on the formation of dwarf galaxies.

Finally, the spherical model adopted in our simulations is unquestionably
oversimplified, as observed outflows such as those in M82 (eg.,
Shopbell \& Bland-Hawthorn 1998) eject matter preferentially along the
spin axis of the galaxy.  Adopting a collimated model would result in
slightly larger accretion rates, as matter would be able to more
easily fall onto the galaxies along the plane of the disk.  In such a
model, the mass shift in Figure \ref{fig:number} would be reduced,
dependent on the unknown beaming angle.  Galaxy suppression by baryonic
stripping, on the other hand, is likely to change little in such a
model, as the spin axes of galaxies are only weakly correlated with
large scale structure, and furthermore only poorly resolved for the
smaller galaxies in our simulations.

\section{Conclusions}

Despite the overwhelming observational evidence for galaxy outflows
at high and low redshift, the impact of this process on 
galaxy formation has been little explored.  While all numerical
galaxy formation simulations carried out to date assume formation
in primordial conditions, exploratory semi-analytical studies
have suggested that many galaxies form under much different
circumstances.  

In this work, we have implemented a simple ``blow-out'' model for
feedback from supernova generated winds, to study the impact of outflows
on the formation of high-redshift dwarf galaxies in a characteristic
volume of the Universe.  While these simulations avoid galaxy
self-annihilation or ``blow-away'' by construction, they have uncovered 
two major mechanism by which outflows impact the formation of dwarf
galaxies.

Firstly, outflowing dwarf galaxies impact their own formation 
by slowing down the further accretion of gas.
This causes an overall decrease of about 50\% in the total
gas mass accreted by these objects which is 
roughly constant over
the mass range studied in our simulations.

In addition to this mechanism, and perhaps more interestingly, our SPH
simulations have also uncovered a significant number of would-be dwarves
whose formation has been suppressed by baryonic stripping by
neighboring objects.  In these cases, the winds from nearby objects
reach the overdense region sufficiently early to eject the gas from 
the collapsing dark-matter perturbation.   This results in a large
fraction of $\sim 10^9 M_\odot$ empty halos in which the dark matter
has virialized but no baryonic gas has collapsed.  

While our simulations
are only able to identify this suppression at high redshift, this
mechanism is suggestive of the discrepancy between the number of 
observed Milky-Way satellites
and predictions from CDM models that do not include outflows
(Klypin et al.\ 1999; Moore et al.\ 1999).  The dark-halos
generated in this scheme also provide a natural explanation
for the existence of massive yet extremely gas-poor extra-galactic
High-Velocity Clouds (see eg.\ Blitz et al.\ 1999),
as well as the observed lack of low-mass galaxies in the field
relative to dense clusters (eg.\ Phillips et al.\ 1998).
Note that as pointed out by Trentham, M\"oller, and Ramirez-Ruiz (2000)
the existence of a large number of empty halos does not depend 
on the detailed predictions of the CDM model but follows simply from 
the assumption that the mass function of dark matter halos
does not vary with environment, thus suggesting that the observed 
enviromentally dependent suppression of dwarf formation is
due to astrophysical considerations rather than some unknown
properties of the dark matter or the primordial power spectrum
(eg.\ Kamionkowski \& Liddle 2000; Spergel \& Steinhardt 2001).

Our results rely on a simple model of outflow formation
in a limited volume of the Universe at redshifts above 6, and thus
can not be directly compared with observations.
Nevertheless the lesson to be drawn from these simulations is clear.
Just as in the case of galaxy clusters, outflows  profoundly 
affect the formation and distribution of dwarf galaxies.
These effects are not confined simply to the
properties of the outflowing galaxies themselves but the 
formation of objects nearby, and this interaction between
neighbors plays an important role in  the history of the formation 
of galaxies. While the assumption of primordial conditions is a good 
first step in the study of this problem then, we will never fully 
understand galaxy  formation until we have first understood  
mechanical feedback from galaxy outflows.

\acknowledgments
We would like to thank Tom Broadhurst and Andrea Ferrara
for helpful comments and discussions. 
ES has been supported in part by an NSF fellowship.
This project was supported by NSF KDI Grant 9872979.

\fontsize{9}{11pt}\selectfont

\end{document}